\documentclass[showpacs,twocolumn,floatfix,prl]{revtex4}
\usepackage{graphicx}
\usepackage{amsfonts}
\usepackage{amsmath}
\usepackage{amssymb}

\renewcommand{\section}[1]{{\par\it #1.---}\ignorespaces}

\begin{document}
\title{ac-driven atomic  quantum motor}
\author{A. V. Ponomarev}
\author{S. Denisov}
\author{P. H\"anggi}
\affiliation{Institute of Physics, University of Augsburg,
Universit\"atstr.~1, D-86159 Augsburg}
\date{\today}
\begin{abstract}
We invent an ac-driven quantum motor consisting of two different,
interacting ultracold atoms placed into a ring-shaped optical
lattice and submerged in a pulsating magnetic field. While the first
atom carries a current, the second one serves as a quantum starter.
For fixed zero-momentum initial conditions the asymptotic carrier velocity
converges to a unique non-zero value. We also demonstrate that
this quantum motor performs work against a constant load.

\end{abstract}


\maketitle

Linear or rotational motion presents the basic working principle powering all sorts of machines.
For nearly two centuries, since the invention of the first
electrical motor \cite{first}, the ever continuing miniaturization of
devices has profound consequences for several branches of science,
industry, and everyday life. This process has already passed the scale of
micrometers  \cite{MEMS} and has entered the realm of the world of nanoscale
\cite{nano1}. Bioinspired devices such as chemical or light driven synthetic molecular
motors identify just one of those recent successes \cite{mol_motors}. While the operational description of such
molecular motors mainly rests on classical concepts, much less is known for operational schemes that are
{\it fully} quantum mechanical in nature. An ideal resource for the latter possibility  is the dynamics of cold atoms
that are positioned in optical potentials \cite{ober}.

With this work, we put forward a setup for a quantum motor which
consists of two species of interacting, distinguishable  quantum particles that are
loaded into a ring-shaped optical potential. The blueprint for
such an underlying ring-shaped one-dimensional optical lattice has been
proposed recently \cite{amico} and a first  experimental realization
has been reported in \cite{ring_exp}. Here, we employ this setup to
devise an  engine which works as a genuine ac-driven quantum motor.

\section{ac-quantum motor}
Figure 1 outlines our device. The ring-shaped optical potential,
which results either from the interference of a Laguerre-Gauss (LG)
laser beam with a plane wave \cite{amico} or, alternatively, of two
collinear LG beams with different frequencies \cite{ring_exp} is capable of trapping
two interacting atoms. One of the atoms, termed ``carrier'', $c$, is
driven by an external field, while the other atom, termed
``starter'',  $s$, interacts locally via elastic s-wave collisions
with the ``carrier'', but remains unaffected by the driving field
\cite{spinless}. Two possible setups that come to mind are
 $(\mathrm{i})$ A neutral ``starter'' and an ionized ``carrier'',
a suitable driving field can be implemented in a way
typically done for  electrons placed in a conducting
ring, i.e., by a time-dependent magnetic flux threading the lattice
\cite{hub_rings}. $(\mathrm{ii})$ A spinless ``starter'' and a
``carrier''-atom with a non-zero spin which is driven by
a time-dependent cone-shaped magnetic field of an Ioffe-Pritchard trap \cite{Kett1, amico}.

We next assume that both atoms are loaded into the lowest energyband
of a deep, ring-shaped optical potential with $L$ lattice sites and 
the lattice constant $d$. The time-dependent homogeneous vector 
potential $\tilde{A}(t)$ does not induce any appreciable transitions 
between the ground band and the excited band(s).

\begin{figure}[t]
\center
\includegraphics[width=0.36\textwidth]{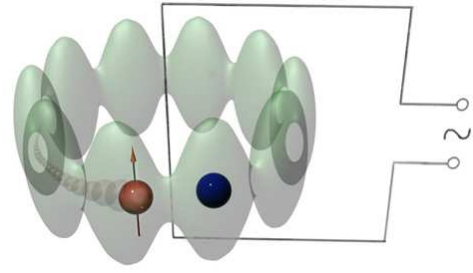}
\caption[Ring-shaped optical lattice] {(color online) Atomic quantum
motor: Two different ultracold  atoms are loaded into a ring-shaped
optical lattice. Both atoms  interact locally with each other, while
only one carrier (the one with an arrow) is magnetically powered.}
\label{fig1}
\end{figure}

The ac-driven, total Hamiltonian $H_{\rm tot}$ of the motor
\begin{equation}
\label{eq:hamiltonian_total} H_{\rm tot} = H_{\rm
c}(t) + H_{\rm s} + H_{\rm int}\,,
\end{equation}
is composed of the time-dependent Hamiltonian $H_{\rm_c}(t)$ for the ``carrier''
\begin{equation}
\label{eq:hamiltonian_fermion} H_{\rm c}(t) = -\frac{J_{\rm
c}}{2}\left(\sum_{l_c=1}^Le^{i\tilde{A}(t)}|l_c+1\rangle_{\rm}\langle
l_c|_{\rm} + {\rm H.c.}\right)\otimes\mathbf{1}_s\,,
\end{equation}
and for the ``starter'' $H_{\rm s}$, respectively, i.e.,
\begin{equation}
\label{eq:hamiltonian_boson} H_{\rm s}= -\frac{J_{\rm
s}}{2}\left(\sum_{l_s=1}^L|l_s+1\rangle_{\rm}\langle l_s|_{\rm} +
{\rm H.c.}\right)\otimes\mathbf{1}_c\,.
\end{equation}
Here, $J_c$ and $J_{s}$ are the corresponding hopping strengths
which are functions of the atom masses and the optical potential depth
\cite{estimate}. The salient carrier-starter (on-site)-interaction reads
\begin{equation}
\label{eq:interaction} H_{\rm int}=W \sum_{l_c,l_s=1}^L \delta_{l_c,l_s}|l_c\rangle_{\rm}\langle l_c|_{\rm}\otimes |l_s\rangle_{\rm}\langle l_s|_{\rm} \,,
\end{equation}
where $W$ denotes the interaction strength. Throughout the
remaining we use periodic boundary conditions; i.e., $|L+1\rangle =
|1\rangle$. The full system Hilbert space is spanned by the direct products 
of single particle Wannier states $|l_c\rangle \otimes |l_s\rangle$, 
with the dimension being $\mathcal{N}=L^2$. The scale of the motor 
current will be measured in units of the maximal group velocity 
$\upsilon_0= J_c d/\hbar$.

The driving of the ac-atomic quantum motor  is switched on at
the time instant $t_0$, so that the vector potential assumes the form
\begin{equation}
\label{eq:switching_time} \tilde{A}(t) = {\mathit\chi(t-t_0)}A(t)\,.
\end{equation}
where ${\mathit\chi(t-t_0)}$ is the step function, and $A(t)$ is
defined on the entire time axis, $t \in (-\infty, +\infty)$.

\section{dc-quantum current}
The mean ``carrier'' current is given as the speed of the motor by
using the velocity operator: $\hat
{\upsilon}_{c}(t)=i/\hbar\left[H_{\rm tot}(t),\hat x_c\right]$. With
$\hat x_c = \sum_ll_c|l_c\rangle_{\rm}\langle l_c|_{\rm}$, one finds
$\hat {\upsilon}_{c}(t)=-i(\upsilon_0/2) \left(\sum_{l_c=1}^L
e^{iA(t)}|l_c+1\rangle_{\rm}\langle l_c|_{\rm}\right.$ $-\left.{\rm
H.c.} \right)\otimes\mathbf{1}_s$. In the quasimomentum
representation with $|\kappa\rangle_{\rm} = \sum_{n=1}^L\exp(i\kappa
n)|n\rangle_{\rm}$, its quantum expectation
${\upsilon}_{c}(t;t_0)=\langle\psi(t)|\hat
\upsilon_c(t)|\psi(t)\rangle$ reads
\begin{equation}
\label{eq:k_current} {\upsilon}_{c}(t;t_0) = \upsilon_0
\sum_{l_c=1}^{L} \rho_{\kappa_{l_c}}(t;t_0)
\sin\left(\kappa_{l_c}+\tilde{A}(t)\right)\,,
\end{equation}
wherein $\kappa_{l_c} = 2\pi l_c/L$ is the single particle
quasimomentum and where we indicated its parametric dependence on
the start time $t_0$. Further,
$\rho_{\kappa_{l_c}}(t; t_0)  = \sum_{l_s}|\langle\psi(t)|\kappa_{l_s}\rangle\otimes|\kappa_{l_c}\rangle|^2$, with $\kappa_{l_s} = 2\pi l_s/L$, 
is the quasimomentum distribution for the carrier.

The steady state regime of the motor can be characterized by the dc-component of the averaged velocity
\begin{eqnarray}
\label{eq:current_average_d} \upsilon_{c}(t_0) := {\rm
lim}_{t\rightarrow\infty}\frac{1}{t}\int_0^t\upsilon_{c}(t' ;
t_0)dt'.
\end{eqnarray}

In the absence of the interaction between the particles, i.e., $W = 0$,
and an initial preparation with
localized carriers that start out with zero velocity
the motor set-up cannot even support a transient directed current. In fact, this result
holds for \textit{any} shape of the vector potential $A(t)$ \cite{Hanggi1}.
This situation thus mimics  a single-phase
ac-motor: a periodically pulsating magnetic field would fail to put
a rotor from rest into rotation, unless one applies an initial ``push"
via a starter mechanism \cite{motor}. In our setup, the role of the
quantum starter is taken over by the non-vanishing
interaction $W$ with the second particle.

Nonetheless, even with nonvanishing interaction, $|W| > 0$, there exists no evident
procedure for setting the motor into rotation. A seemingly obvious solution --
application of a constant carrier-bias  -- cannot resolve the task. This is true because
the corresponding vector potential, $A_{B}(t)=\omega_B t$, induces Bloch oscillations only \cite{Ponomarev}. In
distinct contrast, we shall employ an unbiased time-dependent vector
potential possessing a zero dc-component, $A(t+T)=A(t)$, i.e.,
$\int_{0}^{T}A(\tau)d\tau = 0$.

For zero-momentum initial conditions, the unbiased monochromatic
ac-force, $A(t)=A \sin(\omega t)$, would launch -- with equal
probabilities -- the system either into a clockwise (rightward) or a
counterclockwise rotation (leftward motion) \cite{quantum}. Thus,
the {\it modus operandi} as a motor requires a symmetry-breaking
driving field, realized here with the harmonic mixing signal,
\begin{equation}
\label{eq:driving_vector_potential} A(t) = A_1\sin(\omega
 t)+A_2\sin(2\omega t+\Theta)\,,
\end{equation}
where $\Theta$ denotes the crucial symmetry-breaking phase shift.
The input (\ref{eq:driving_vector_potential}) knowingly may induce a
non-vanishing nonlinear response, the so-called \textit{ratchet
effect} \cite{quantum, ratchet, Renzoni}.

\section{Quantum current in terms of Floquet states}
The  dynamics at times $t > t_0$ of the time-periodic Hamiltonian
(\ref{eq:hamiltonian_total}) can be analyzed by using the Floquet
formalism \cite{Hanggi2}. The solution of the eigenproblem:
$U(t,t_0)|\phi_n(t;t_0;k)\rangle =
\exp\left(-\frac{i}{\hbar}\epsilon_nt\right)|\phi_n(t;t_0;k)\rangle$,
with the propagator $U(t,t_0) = {\cal T}\exp
\left(-\frac{i}{\hbar}\int_{t_0}^t H_{\rm tot}(\tau)\right)d\tau$
(${\cal T}$ denotes the time ordering), provides the set of Floquet
states, being time-periodic, i.e., with $T= 2\pi/\omega$ being the
driving period, $|\phi_n (t+T; t_0;k)\rangle = |\phi_n(t;
t_0;k)\rangle$. Here, $k=\sum_{l,m}
\langle\phi_l|\kappa_l\rangle_{\rm s}\otimes|\kappa_m\rangle_{\rm
c}$ is the {\it total} quasimomentum of the Floquet state. Because of
the discrete translation invariance of the system, the total
quasimomentum is conserved during the time evolution, thus serving
as a quantum number. Since $H_{\rm tot}$ is a function of the time
difference $t-t_0$ only, the quasienergies $\epsilon_n$ are
independent of $t_0$, and the Floquet states for different
start times $t_0$ are related by $|\phi_n(t; t_0;k)\rangle =|\phi_n(t-t_0; 0; k)\rangle$.

Using this relation we next decompose $\psi(t_0)$ in the complete
basis of Floquet states:
$|\psi(t_0)\rangle = \sum_{n=1}^\mathcal{N} c_n |\phi_n (0; t_0;
k)\rangle$ $=\sum_{n=1}^\mathcal{N} c_n |\phi_n (T-t_0; 0;
k)\rangle$. In the absence of the driving, $A(t)\equiv 0$, the
motor set-up (\ref{eq:hamiltonian_total}-\ref{eq:hamiltonian_boson}) possesses the
continuous translational symmetry in time. Thus, the expansion coefficients of the initial
wave-function $\psi(t_0)$ in the system eigenbasis knowingly do not depend
on time. On the contrary, eigenstates of a periodically driven
system -- the Floquet states --  evolve in time, being locked by the
external ac-field. The expansion of an initial wave-function over
the Floquet eigenbasis depends on the start time $t_0$
(\ref{eq:switching_time}), which determines the phase of the driving
ac-field \cite{quantum}: $c_n=c_n(t_0)=\langle \phi_n (T-t_0,0;
k)|\psi(t_0)\rangle$. Substitution of the above decomposition into
(\ref{eq:current_average_d}) yields the result
\begin{eqnarray}
\label{eq:current_average} \upsilon_{\rm c}(t_0)
=\sum_{n=1}^{\mathcal{N}} \overline{\upsilon}_n\, |c_n(t_0)|^2 \;,\;
\overline{\upsilon}_n = \frac{1}{T} \int_0^{T} \upsilon_n(\tau)
d\tau\,.\;
\end{eqnarray}
Here, $\overline{\upsilon}_n$ denotes the quantum average velocity of the
$n$-th Floquet state (\ref{eq:k_current}). Because the Floquet states
are periodic functions of the time difference $\tau=t-t_0$ only, the
velocities $\overline{\upsilon}_n$ do not depend on $t_0$, and the
dependence of the generated dc-current on the $t_0$ solely stems
from the coefficients~$c_n(t_0)$. Since the system evolution is fully quantum coherent,
i.e. there is no memory
erasing induced by an environment, the asymptotic current maintains the memory of the
initial condition as encoded in the coefficients $c_n(t_0)$ \cite{quantum}.

\section{Input/Output characteristics}
The question is now, how can we control the motor? To answer this
question, we used the symmetry analysis \cite{quantum} which allows
us to predict an appearance of a certain dc-current. Combining
time-reversal operation and the complex conjugation applied to
(\ref{eq:k_current}) with $A(t)$ in the form of
(\ref{eq:driving_vector_potential}), one can prove the (anti-)
symmetric dependence of $\overline{\upsilon}_n$ on $\Theta$ for the
Floquet states with $k=0$: $\overline{\upsilon}_n(\pi-\Theta) =
\overline{\upsilon}_n(\Theta)$, $\overline{\upsilon}_n(-\Theta) =
-\overline{\upsilon}_n(\Theta)$. Thus, the Floquet states with $k=0$
possess zero mean velocities at $\Theta=0, \pi$. Furthermore,  using
a similar reasoning, one finds that the set of Floquet states with
nonzero $k$ can be ordered by the parity relation, which links
eigenstates with opposite quasimomenta, $\phi_n(t; t_0; -k;\Theta) =
\phi_m(T-t; t_0; k; -\Theta)$, yielding $\bar{\upsilon}_n =
-\bar{\upsilon}_m$.  This implies that for a symmetric (in $k$)
initial state  and $\Theta=0,\pi$, the contributions to the
dc-current of Floquet states with opposite quasimomenta eliminate
each other. The same holds true for a monochromatic driving
(\ref{eq:driving_vector_potential}), with $A_2=0$ \cite{quantum}.
Shifting $\Theta$ away from  $0,\pm\pi$ causes the decisive symmetry
breaking and leads to the desymmetrization of the Floquet states
with $k=0$ and consequently will violate the parity between states
with opposite signs of $k$.

\begin{figure}[t]
\center
\includegraphics[width=0.42\textwidth]{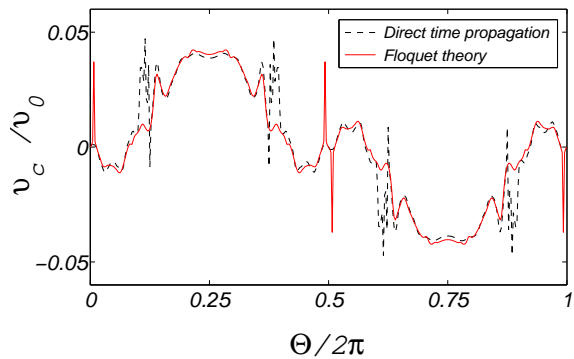}
\caption {(color online) Averaged motor velocity in (\ref{eq:MC})
(in units of the maximal group velocity $\upsilon_0 = J_{\rm c}d/\hbar$) as
a function of the phase shift $\Theta$ in
(\ref{eq:driving_vector_potential}) for $L=16$. The ($t_0$)-averaged
velocity (\ref{eq:current_average_d}) obtained by the direct time
propagation of the initial state up to $200T$ (dashed line) is
compared to the asymptotic dependence given by the Floquet approach
(\ref{eq:current_average}) (red solid line).  The parameters are
$\hbar\omega = 0.1\times J_{\rm c}$, $A_1 = 0.5$, $A_2 = 0.25$, $W =
0.2 J_{\rm c}$, $J_{\rm s} = J_{\rm c} = J$.} \label{fig2}
\end{figure}

The motor speed depends on the initial conditions, which define the
contributions of different Floquet states to the carrier velocity
(\ref{eq:current_average}). We restrict our analysis to the initial
state $\psi(t_0)=L^{-1/2}|l_c\rangle\otimes\sum_{l_s}|l_s\rangle$,
$l_c=1,...,L$, in the form of the localized carrier (at $l_c$) and
the uniformly ``smeared", delocalized starter. Both  particles have
zero velocities at $t=t_0$. The asymptotic velocity typically
exhibits a strong dependence on $t_0$ \cite{quantum}. We first
discuss the results obtained after averaging over $t_0$, thus
assigning a unique motor velocity value,
\begin{equation}
\label{eq:MC}
\upsilon_c = \langle \upsilon_c(t_0) \rangle_{t_0}
= 1/T \int_{t_{0}}^{T+t_{0}}\upsilon_c(t_0)dt_0 \;,
\end{equation}
for fixed system parameters.

Figure \ref{fig2} depicts the dependence of the average motor
velocity on $\Theta$. The results obtained by  direct time
propagation   of the initial state and averaged over $t_0$ (dashed
line) are superimposed by those calculated via the Floquet formalism
(\ref{eq:current_average}) (solid line). The agreement between the
two curves is satisfactory but not perfect: This is true because of the
sharp peaks on the asymptotic current (\ref{eq:current_average}).

\begin{figure}[t]
\center
\includegraphics[width=0.35\textwidth]{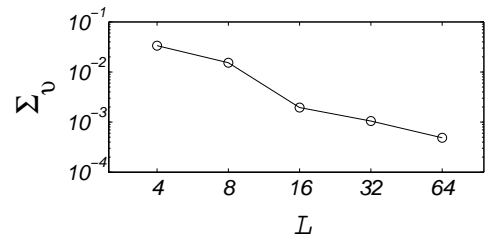}
\caption[Dispersion of the motor velocity.] {Dispersion of the motor
velocity (\ref{eq:dispersion}) {\it versus} the number of lattice sites
$L$. Here $\Theta=\pi/2$, and the other parameters are the same as
in Fig. \ref{fig2}.} \label{fig3}
\end{figure}

These  peaks can be associated with \textit{avoided crossings}
between two quasienergy levels \cite{quantum}. These avoided
crossings cause a strong current enhancement if one of the
interacting, and transporting eigenstate overlaps significantly with
an initial, nontransporting state of the motor. Note also that
a very narrow avoided crossing requires a very large evolution time to become
resolved, i.e., $t_{obs} \sim \hbar/|\epsilon_{\alpha} -
\epsilon_{\beta}|$ \cite{quantum}. Our chosen  evolution time $t = 200T$ is
not large enough to clearly resolve the distinct resonances depicted in Fig. \ref{fig2}.

We further detect that the  dependence of the motor velocity
$v_c(t_0)$ in (\ref{eq:current_average_d}) on $t_0$ increasingly
disappears upon increasing the size $L$ (not shown). To provide a
quantitative estimate, we evaluated the dispersion of the current
(\ref{eq:current_average}) with respect to $t_0$, i.e.,
\begin{equation}
\label{eq:dispersion} \Sigma_\upsilon =
\sqrt{\left\langle\upsilon_{c}(t_0)^2-\langle\upsilon_{c}(t_0)\rangle_{t_0}^2
\right \rangle}_{t_0}\,.
\end{equation}
As shown in Fig.~\ref{fig3} this dispersion decays with increasing
size $L$, being rather faint for $L\gtrsim 16$.  For sizes $L
\gtrsim 16$ the carrier assumes an asymptotic velocity that
essentially is independent of the initial start time $t_0$. This
effect is caused by the presence of the starter:  The carrier
velocity is obtained as the trace over the part of the total system
Hilbert space, associated with the starter. The starter dynamics
mimics a dissipative, finite heat bath for the carrier dynamics
whose effectiveness increases with both, the (i) the dimension of
the starter subspace, i. e. the size  $L$, and (ii) the strength of
the interaction $W$.

\begin{figure}[t]
\center
\includegraphics[width=0.42\textwidth]{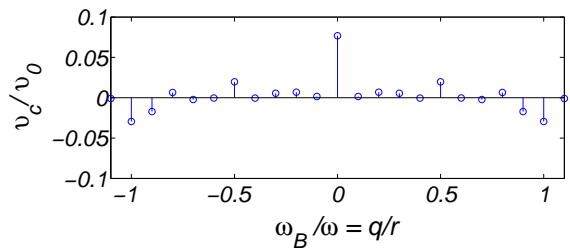}
\caption {The average motor velocity for
``resonance'' driving and an initial condition ``localized
carrier and delocalized starter" ($r=10$). The parameters are $W=0.2
J_c$, $J_s = J_c= J$, $\hbar\omega = 0.1  J_c$, $A_1=0.5$,
$A_2=0.25$, $\Theta=\pi/2$, and $L=4$.} \label{fig4}
\end{figure}

\section{Load characteristics} The analysis based on Eqs.~(\ref{eq:hamiltonian_total}
- \ref{eq:hamiltonian_boson}, \ref{eq:driving_vector_potential}) has
been for a free rotator. In order to qualify for a genuine motor
device, the engine must be able to operate under an applied load.
The load is introduced as the bias $\omega_Bt$, being added to the
vector potential $\tilde{A}(t)$. All the information about transport
properties can be extracted by using again the Floquet formalism,
provided that the ac-driving and the Bloch frequencies are mutually
in resonance \cite{kolovsky}, i.e. $q\omega = r \omega_B$, where $r$
and $q$ are co-prime integers. Figure 4 shows the dependence of the
asymptotic motor speed for different bias values. There are two
remarkable features. First, the spectrum of velocities is symmetric
around $\omega_B=0$. This follows because of the specific choice of the
phase shift at $\Theta=\pi/2$. Second, while some regimes provide a
transport velocity along the bias, others correspond to the up-hill
motion, against the bias. Therefore, a stationary transport in
either direction is feasible. The load characteristics exhibits a
discontinuous, fractal structure and, in distinct contrast to the
classical case \cite{kostur}, it cannot be approximated by a smooth
curve. This is a direct consequence of the above mentioned resonance
condition.

\section{Experimental realizations} For an experimental realization 
of this quantum atom motor the
following features should be respected: (i) In the case of the setup
``carrier with a spin/spinless starter", the carrier should assume a
magnetic number $m_F \geq 2$ \cite{amico}, to efficiently  induce
the ac-field amplitudes $A_{1,2}$. (ii) Because in the tight-binding
approximation the maximal amplitude of the tunneling is limited from
above, $J_c \lesssim J_{max}=0.13 E_0$, for $^6\rm Li$ atom, the
lattice spacing $d\sim 10\mu m$ \cite{ring_exp}, and $\hbar \omega =
0.1 J_c$ (used in the calculations), the driving frequency $\omega$
should be less than $2$Hz. Then, the time required to launch the
motor (i.e., to approach the asymptotic velocity value) is around a
minute. Further focusing of the laser beam can decrease the lattice
constant $d$, thereby decreasing the launch time to experimentally
accessible coherence times around $10$ seconds \cite{gustavsson}.

\section{Conclusions}
We studied  a quantum ac-motor made of the two species of ultracold
interacting atoms, i.e. a ``carrier" and a ``starter", moving in a
ring-shaped trapping potential.  For zero-momentum initial
conditions the  asymptotic carrier velocity loses its dependence
on the switch-on time $t_0$ of the ac-drive upon increasing the 
lattice size $L$.  A natural question that arises is, What about 
the averaged starter velocity $v_s$?  
We find that the latter sensitively depends on the system parameters:
It can either be very small compared to the carrier velocity or
also larger than $v_c$. In short, the starter can move
co-directionally or contradirectionally to the carrier motion.
Finally, an extension of our motor setup to  several interacting
bosons (i.e. a finite bosonic ``heat bath'') presents an intriguing
perspective.

We acknowledge I. V. Ponomarev for providing us with the
illustration shown in Fig.\ref{fig1}. This work was supported by the
DFG through grant HA1517/31-1 and by the German Excellence
Initiative ``Nanosystems Initiative Munich (NIM)".


\begin{thebibliography}{1000}
\bibitem{estimate} Here, for simplicity, we have assumed the limit 
of a deep periodic potential, also refereed to as the tight-binding model. 
This limit is well justified for a potential amplitude of $V_{0} \gtrsim 5E_0$ 
($E_0=\hbar^2\pi^2/2Md^2$ is the ``recoil'' energy, $M$ is the mass of the atom), 
where the higher-order tunneling amplitude is less than $10\%$ of the leading first-order 
(i.e., the next neighbour) tunneling amplitude \cite{ober}.
\end{thebibliography}
\end{document}